\documentclass[aps,twocolumn,groupedaddress,superscriptaddress,floatfix,amsmath,amssymb,prb]{revtex4-1}
\usepackage{amsmath}
\usepackage{amssymb}
\usepackage{graphicx}
\usepackage{textcomp}
\usepackage{bm}

\usepackage{booktabs}
\usepackage{siunitx}

\usepackage[usenames,dvipsnames]{color}	

\begin{document}
\title{Digital Quantum Simulations of Spin Models on \\ Hybrid Platform and Near-Term Quantum Processors}


\author{F. Tacchino}
\email{francesco.tacchino01@universitadipavia.it}    
\affiliation{Dipartimento di Fisica, Universit\`a di Pavia, via Bassi 6, I-27100 Pavia, Italy}
\author{A. Chiesa}
\affiliation{Dipartimento di Scienze Matematiche, Fisiche e Informatiche, Universit\`a di Parma, I-43124 Parma, Italy}
\author{M. D. LaHaye}
\affiliation{Department of Physics, Syracuse University, Syracuse NY 13244-1130, USA}
\author{I. Tavernelli}
\affiliation{IBM Research, Zurich Research Laboratory, Zurich, Switzerland}
\author{S. Carretta}
\affiliation{Dipartimento di Scienze Matematiche, Fisiche e Informatiche, Universit\`a di Parma, I-43124 Parma, Italy}
\author{D. Gerace}
\email{dario.gerace@unipv.it}    
\affiliation{Dipartimento di Fisica, Universit\`a di Pavia, via Bassi 6, I-27100 Pavia, Italy}

\begin{abstract} 
We review a recent theoretical proposal for a universal quantum computing platform based on tunable nonlinear electromechanical nano-oscillators, in which qubits are encoded in the anharmonic vibrational modes of mechanical resonators coupled to a superconducting circuitry. 
The digital quantum simulation of spin-type model Hamiltonians, such as the Ising model in a transverse field, could be performed with very high fidelities on such a prospective platform.
Here we challenge our proposed simulator with the actual IBM-Q quantum processor available on cloud. We show that such state-of-art implementation of a quantum computer, based on transmon qubits and superconducting technology, is able to perform digital quantum simulations. However, encoding the qubits in mechanical degrees of freedom would allow to outperform the current implementations in terms of fidelity and scalability of the quantum simulation.
\end{abstract}

\maketitle



\textit{Introduction}.
Digital quantum simulators are among the most appealing applications of a quantum computer \cite{Georgescu2014,Laflamme}. In principle, any model that can be mapped onto a spin-type Hamiltonian can be encoded in a digital quantum simulator. Then, its time evolution can be solved with arbitrary precision, thus overcoming the unavoidable exponential scaling of computational resources that is inherent to quantum manybody physics \cite{Lloyd1996}. Several theoretical proposals and proof-of-principle experiments on different platforms have already demonstrated the validity and huge prospective potentialities of such techniques \cite{Schindler2013,LasHeras2014,Barends2015,Salathe2015,Santini2011}. 
In particular, superconducting qubits are nowadays regarded as the most promising candidates towards the practical realization of an actual platform for universal quantum computation in the long run \cite{Gambetta2017}, as well as for applications in noisy intermediate-scale quantum devices (NISQ) in the next few years \cite{PreskillNISQ,Kandala2017}. However, it is still not clear whether this will be the ultimate technological choice for all possible applications: in this respect, the rapidly growing field of hybrid quantum technologies offers a plethora of alternative candidates that are designed to enhance performances through the integration of different quantum systems, including spin ensembles, opto- and electromechanical devices, and photons \cite{Kurizki2015}. 
Diverse hybrid solutions on superconducting platforms were already proposed for applications to digital quantum simulations \cite{Carretta2013,Chiesa2014,Chiesa2015,Chiesa2016}.

Along these lines, we hereby present a brief review of our recent theoretical proposal for a universal, scalable, and integrated quantum computing platform based on tunable nonlinear electromechanical nano-oscillators \cite{Tacchino2018}. In a minimal version of such architecture, qubits could be encoded in the anharmonic vibrational modes of nanomechanical resonators coupled to a superconducting nanocircuitry. Practical realizations of such qubits can be envisioned as suspended nanotubes, two-dimensional nanomembranes (e.g. graphene sheets), or cantilevers, all showing very promising quality factors and coherence times up to several milliseconds. In the proposed platform, single-qubit rotations can be implemented by using external static and modulated electric fields acting locally on a single nanoresonator, while two-qubits gates would be efficiently realized by mediating their effective coupling through virtual fluctuations of an intermediate superconducting artificial atom, such as a transmon. As it was shown in the original Ref.\ \onlinecite{Tacchino2018}, and at difference with the use of the superconducting elements as qubits, the transmon coherence time ($T_2$) becomes essentially irrelevant in our proposed platform, which helps envisioning considerable prospects for scalability.

As explicit proof-of-principle examples of the electromechanical quantum simulator theoretical performances, we show numerical results for the digital quantum simulation of the time evolution of few spin model Hamiltonians. Moreover, we challenged our proposed simulator with an existing NISQ one, i.e. the IBM Q quantum processor freely available for cloud quantum computation \cite{ibmq}. Interestingly, such state-of-art implementation of an actual quantum computer, whose hardware employs purely superconducting qubits in a microwave nanocircuit cooled to few mK in a dilution fridge, is potentially able to perform digital quantum simulations of a few targeted models of interest in condensed matter physics. We have used the results obtained on these models to benchmark our theoretically proposed platform.

\begin{figure*}[t]
\centering
\includegraphics[width=0.8\textwidth]{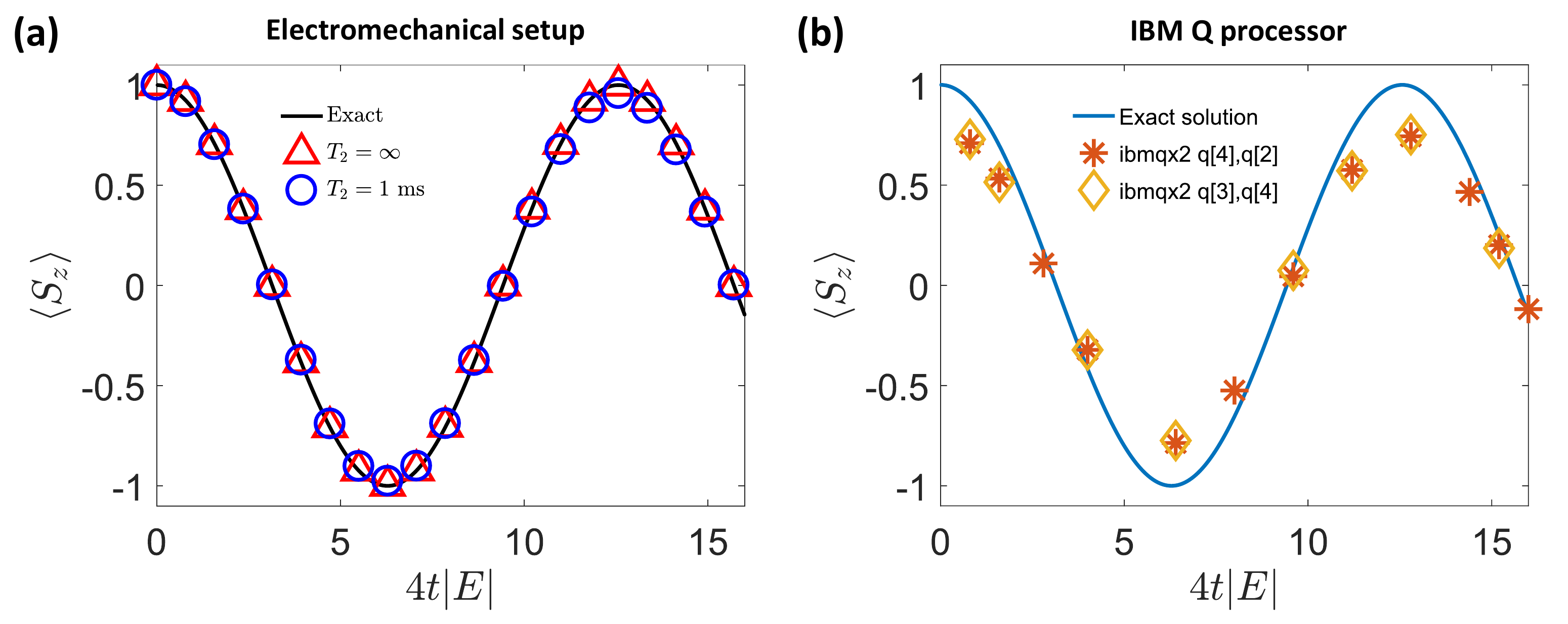}
\caption{Digital quantum simulation of the tunnelling of the total magnetization in a spin-1 Hamiltonian. (\textbf{a}) Numerical results for the proposed electromechanical set-up, for different values of the NRs $T_2$ time. (\textbf{b}) Experimental data from the IBM Q ibmqx2-Yorktown real backend, for different choices of the pair of qubits selected on the chip.}
\label{fig:S1}
\end{figure*}

\textit{Electromechanical qubits}.
The fundamental unit of our proposed architecture is given by a pair of electromechanical nanoresonators (NRs) coupled through a nonlinear circuit element, e.g.\ a transmon. This elementary building block can be described with the Hamiltonian

\begin{equation}
H_{0}= \sum_{i=1}^2 \left[ \omega_i b^\dagger_i b_i + H_{nl,i} \right] + \frac{\Omega} {2}\sigma_z \, ,
\label{eq:H0}
\end{equation}

where $b_i$ ($b_i^\dagger$) represent bosonic annihilation (creation) operators, and $H_{nl,i}$ explicitly introduces the required anharmonicity to isolate the two lowest energy levels of the NRs, where the qubits are encoded. For simplicity, the transmon is treated as a pure quantum two-level system with $\sigma_\alpha$ ($\alpha=x,y,z$) representing Pauli matrices. The interaction between mechanical oscillators and the transmon is modeled as 
\begin{equation}
H_{int} = \sum_{i=1}^2 g_{i} \left(b_i+b_i^\dagger\right)\sigma_x.
\end{equation}
In the strongly dispersive limit, where the frequency of the NRs (typically in the MHz range) is far from the transition frequency of the transmon (typically a few GHz), the effective transmon-mediated interaction Hamiltonian between the two electromechanical qubits, when restricted to the computational basis, takes the form of an effective XY exchange coupling $H_{eff}\propto \sigma_{x,1}\sigma_{x,2} + \sigma_{y,1}\sigma_{y,2}$. This provides the basis for a $\sqrt{\mathrm{iSWAP}}$ two-qubit gate which, together with single-qubit rotations, gives a universal set of quantum gates (see Ref. \onlinecite{Tacchino2018} for further details). In all numerical simulations, dissipation and pure dephasing effects on the NRs and the transmon are taken into account by means of the corresponding density matrix master equation \cite{Tacchino2018}.

\begin{figure*}[t]
\centering
\includegraphics[width=0.8\textwidth]{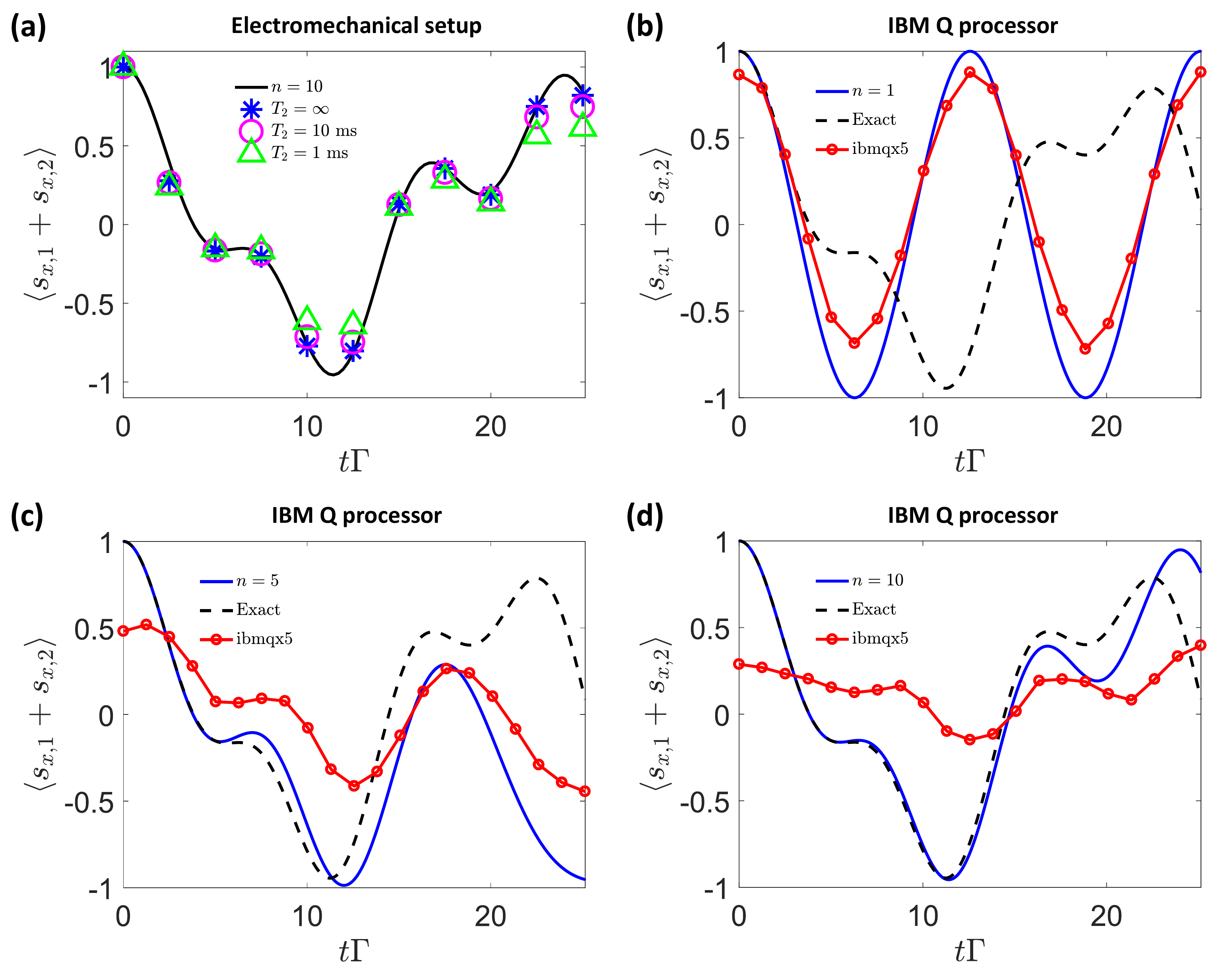}
\caption{Digital quantum simulation of the TIM. (\textbf{a}) Numerical results obtained for the proposed electromechanical set-up, for different values of the NRs $T_2$ and $n=10$ Suzuki-Trotter iterations, compared to the corresponding ideal results (for the same digital decomposition in $n=10$ steps). (\textbf{b-d}) Experimental data from the IBM Q ibmqx5-Rueschlikon real backend, for increasing values of the number of Suzuki-Trotter iterations. The dashed line represents the exact result with no digital error.}
\label{fig:TIM}
\end{figure*}   

\textit{Digital quantum simulations}.
Calculating the dynamics of a given target Hamiltonian, $\mathcal{H}$, is a computationally hard task for classical computers when $\mathcal{H}$ models a system with a large number of degrees of freedom, due to the fast exponential scaling of the Hilbert space dimension. However, the problem was proven to be efficiently solvable on a quantum hardware when the interactions between subsystems are local in nature \cite{Lloyd1996}. In principle, any general purpose quantum processor can be used, provided that the target $\mathcal{H}$ is digitally encoded on the qubits register. Indeed, if $\mathcal{H}$ is a sum of local terms, i.e. $\mathcal{H} = \sum_k \mathcal{H}_k$, one only needs to decompose each individual local unitary operator, $U_k(t) = exp(-i\mathcal{H}_k t)$, into a sequence of quantum gates (this is always possible on any platform providing a universal set of gates), and append such sequences one after the other. The target unitary operation can then be approximated within arbitrary digital precision by dividing the total target time, $t$, into smaller steps $t/n$, and by repeating the total sequence $n$ times, in what is also known as Suzuki-Trotter decomposition

\begin{equation}
e^{-i \mathcal{H}t} \simeq \left(\prod_k U_k(t/n) \right)^n
\label{eq:ST}
\end{equation}

\textit{Results}. In this section, we show a comparison between the predicted quantum simulation capabilities of our proposed electromechanical architecture and a state-of-the-art prototype real quantum processor based on superconducting circuits, which is publicly available through the IBM Quantum Experience \cite{ibmq}. All the results for the electromechanical simulator are obtained numerically by solving the corresponding master equation. The latter fully describes the realistic behavior of the proposed hardware set-up, including all the electrical pulse sequences necessary to control the NRs qubits and to implement the required gates, and takes into account different realistic values of the nanomechanical qubits and transmon damping ($T_1$) and coherence ($T_2$) times. Typical values of the NRs and transmon frequencies employed in the numerical simulations are in the range $\omega_i /2\pi = 75-85$ MHz and $\Omega /2\pi = 2.5-10$ GHz, while the NRs-transmon coupling used in the simulations is $g_i = g = 2\pi \cdot 6$ MHz ($i=1,2$). Again, we refer to the original Ref. \onlinecite{Tacchino2018} for further details about the model and its numerical implementation. In addition, experimental data from the real IBM Q backend are obtained by programming the quantum chip on cloud through the Qiskit Python libraries \cite{qiskit}.

Here, we focus our attention on spin-type Hamiltonians. The time evolution induced by single-body terms of the form $\mathcal{H}_\alpha^{(1)} \propto \sigma_\alpha^i$ directly corresponds to single-qubit rotations $R_\alpha^i$. On the other hand, the unitary evolution induced by terms of the form $\mathcal{H}_\alpha^{(2)} \propto \sigma_{\alpha,1} \sigma_{\beta,2}$ can be obtained in general by combining single qubit rotations with two-qubit operations. The latter are implemented differently on different platforms, thus resulting in different decompositions of the target unitary evolution: in the electromechanical set-up, the XY effective interaction term between NRs can be used, while on the IBM Q devices the controlled-$\mathrm{NOT}$ operation is natively available.

As a first example, we show in Fig.~\ref{fig:S1} the quantum simulation of the tunnelling of the total magnetization in a spin-1 Hamiltonian. In general, $S>1/2$ models can be mapped into the state of $2S$ qubits. In the case considered here, the target $S=1$ Hamiltonian $\mathcal{H}_{S1} = D S_z^2 + E (S_x^2 - S_y^2 ) \,$ is mapped into $\tilde{\mathcal{H}}_{S1} = 2D s_{z,1}s_{z,2} + 2E (s_{x,1}s_{x,2}  - s_{y,1} s_{y,2}) \, $ by considering the total spin as a combination of two $1/2$ spins, $S_{\alpha}=s_{\alpha,1}+s_{\alpha,2}$. As shown in the figure, the overall quantum simulation is predicted to work very well on the electromechanical set up, also for realistic values of the NRs $T_2$ time (Fig.~\ref{fig:S1}a). Interestingly, comparable performances can be obtained by simulating the very same model on the real IBM Q hardware. We notice that in this case the digital decomposition of Eq. \eqref{eq:ST} is already exact with $n=1$, as evidenced in Fig.~\ref{fig:S1}b.


To test the quantum simulators on more complex spin models, we have performed the digital quantum simulation of the total magnetization along $x$, i.e. $\langle S_x \rangle=\mathrm{Tr}[\rho(s_{x,1} + s_{x,2})]$, for the transverse field Ising model (TIM) of two 1/2 spins $\mathcal{H}_{TIM} = \Gamma s_{x,1} s_{x,2}  + b (s_{z,1} + s_{z,2})$, where we set $\Gamma = 2 b$. We stress that the digital quantum simulation is much more demanding in this case. Indeed, since the single and two-qubit terms in $\mathcal{H}_{TIM}$ do not commute, the digitalized computation using the Suzuki-Trotter formula of Eq. \eqref{eq:ST} requires at least $n\simeq 10$ in order to get physically meaningful results for $\Gamma t > 20$. Results for this model are shown in Fig.~\ref{fig:TIM}.
Despite the increased length of the sequences of gates that are involved, our proposed implementation on a hybrid platform shows robust predicted performances with very high fidelities (Fig.~\ref{fig:TIM}a). 
At the same time, the publicly available IBM Q hardware is able to correctly reproduce the expected behaviour up to $n\simeq 5$ and $\Gamma t \simeq 18$, while it is pushed close to its state-of-art limitations for digital simulations requiring longer sequences of quantum gates (e.g. $n=10$), as it is evident by comparing the results in Figs.~\ref{fig:TIM}b-d.

\textit{Discussion}.
We have briefly reviewed an original proposal for implementing an electromechanical universal quantum simulator, and compared its potential performances with existing quantum processors based on superconducting technology. On one hand, the results show the robustness and potential strength of a prospective hybrid quantum computing platform. On the other hand, despite the still noisy and imperfect nature of the data that can be obtained on the real IBM Q processors (similar to other state-of-art prototypes), we demonstrated how these near term devices might already offer unprecedented opportunities to develop novel and useful approaches to quantum computing, either in the physical sciences \cite{Kandala2017,Chiesa2018} or in other areas of research \cite{Schuld2017EPL,Otterbach2017,Tacchino2018neuron}.

\vspace{6pt} 

\acknowledgments{We acknowledge use of the IBM Q for this work. The views expressed are those of the authors and do not reflect the official policy or position of IBM or the IBM Q team. We thank Mr.\ M.\ Grossi for useful discussions. 
This work was partly funded by the Italian Ministry of Education and Research (MIUR) through PRIN Project 2015 HYFSRT  ``Quantum Coherence in Nanostructures of Molecular Spin Qubits''.}





\end{document}